\documentclass[12pt,preprint]{aastex}

\usepackage{psfig}
\usepackage{epsfig}

\begin{document}

\title{Redshift-Independent Distances to Type Ia Supernovae}

\shorttitle{Z-free SNe Ia Distances}

\author{Brian J. Barris\altaffilmark{1} and John L. Tonry\altaffilmark{1}}

\altaffiltext{1}{Institute for Astronomy (IfA), University of Hawaii,
2680 Woodlawn Drive, Honolulu, HI 96822; barris@ifa.hawaii.edu,
jt@ifa.hawaii.edu}

\begin{abstract}
\label{batmloop-abstract}

We describe a procedure for accurately determining luminosity
distances to Type Ia supernovae (SNe Ia) without knowledge of
redshift.  This procedure, which may be used as an extension of any of
the various distance determination methods currently in use, is based
on marginalizing over redshift, removing the requirement of knowing
$z$ a priori.  We demonstrate that the Hubble diagram scatter of
distances measured with this technique is approximately equal to that
of distances derived from conventional redshift-specific methods for a
set of 60 nearby SNe Ia.  This indicates that accurate distances for
cosmological SNe Ia may be determined without the requirement of
spectroscopic redshifts, which are typically the limiting factor for
the number of SNe that modern surveys can collect.  Removing this
limitation would greatly increase the number of SNe for which current
and future SN surveys will be able to accurately measure distance.
The method may also be able to be used for high-$z$ SNe Ia to
determine cosmological density parameters without redshift
information.

\end{abstract}


\section{Introduction}
\label{batmloop-intro}

The use of Type Ia supernovae (SNe Ia) as a tool for cosmological
studies became feasible upon the realization that there is a
relationship between luminosity and light-curve shape (Phillips 1993).
With this knowledge and the advent of wide-field CCD arrays which
allow for the efficient discovery of large numbers of SNe Ia,
cosmological investigations into the fundamental composition of the
universe using SNe Ia as probes have revealed the acceleration of the
expansion of the universe (Riess et al. 1998, Perlmutter et al. 1999).

In order to derive distance from observed photometry, numerous factors
which alter the light-curve must be taken into account.  Any distance
measurement technique must calculate the probability of a fit for a SN
light-curve in the parameter space of ($t_0$, $A_V$, $R_D$, $z$, $d$),
where $t_0$=time-of-maximum, $A_V$=extinction, $R_D$=decline rate
(using any of the several current parameterizations of SNe Ia
light-curve shapes---see below), $z$=redshift, and $d$=distance.  We
then either marginalize over these factors or take cuts through
parameter space at specific values in order to obtain a measurement
for $d$, the primary quantity of interest.

The initial discovery of the relationship between light-curve shape
and brightness led to parameterization by the rate of decline in
$B$-band brightness over the 15 days after maximum light ($\Delta
m_{15}$).  A second method, the Multi-wavelength Light-Curve Shape
(MLCS) method (see Riess et al. 1996, 1998) is based on $\Delta$, the
difference in peak brightness between an observed SN and a fiducial
light-curve template.  The ``stretch'' method (Perlmutter et al. 1997)
parameterizes the light-curve by a factor $s$ which broadens or
narrows a template light-curve in order to modify the light-curve
shape.  The Bayesian Adapted Template Match (BATM) method introduced
by Tonry et al. (2003) measures distances through comparison with a
large set of well-observed nearby supernovae rather than a
parametrized template.  These methods all seem to be fundamentally
equivalent for the purposes of measuring accurate distances and
resolving uncertainties between the effects mentioned above.  Another
recently introduced distance measurement technique is CMAGIC (Wang et
al. 2003), which uses the relationship between light-curve shape and
brightness in a more indirect way than the above methods.  It utilizes
an observed linear relationship in color-magnitude space for extended
periods of time after maximum light to estimate distance.  As noted by
Wang et al. (2003), CMAGIC remains under development and a more
thorough demonstration of its utility as a distance estimator is
necessary.

Redshift is an especially problematic parameter.  It broadens the
light-curve by (1+$z$), which affects what we can deduce about
luminosity, and also modifies how the spectral energy distribution
(SED) maps into observed bandpasses, thereby changing the color of the
SN Ia.  It thus exhibits considerable covariance with other fit
parameters, and needs to be known accurately.  Redshift is normally
measured spectroscopically, which confers the added benefit of
confirming that an object is indeed a SN Ia.  However, for large
surveys (see Barris et al. 2004), far more SNe are discovered than can
possibly be observed spectroscopically.  The situation will get worse
with larger surveys such as ESSENCE (Smith et al. 2002), the
Canada-France-Hawaii Legacy Survey
(http://www.cfht.hawaii.edu/Science/CFHLS) and Pan-STARRS (Kaiser \&
Pan-STARRS Team 2002).

There have been several attempts to determine redshift and type for SN
through photometric means.  Barris et al. (2002) concluded that over a
wide range of redshift, SNe Ia lie in a narrow region of
color-magnitude space, indicating it is likely that tests
discriminating between SNe Ia and II may be reliable.  Riess et
al. (2004) constructed cuts based on SEDs of SNe to optimize selection
of high-$z$ candidates for spectroscopy, distinguishing between the
various SN types at high redshifts based on the UV deficit of SNe Ia
relative to core collapse SNe.  In combination with the photometric
colors of the host galaxy, they were also able to make a rough
estimate of the redshift of the SNe.  Barris et al. (2004) used a
similar, although simplified, comparison to argue for the
identification of several SNe as Type Ia, despite lacking spectral
confirmation.

Currently it is not apparent that photometric observations are
sufficient to accurately determine the redshift and/or distances of
SNe Ia.  Comparisons with SEDs of SNe Ia may allow constraints on
(1+$z$) on the order of $\approx$0.1, but this is a poor constraint on
$z$ for $z<0.5$.  Without knowledge of redshift, the standard methods
for determining distance cannot be meaningfully applied, since
observations must be corrected for cosmological effects such as
time-dilation and K-corrections before allowing accurate measurements
of other parameters.  However, often the quantity of interest is
distance.  If $d$ may be accurately measured despite a poor constraint
on $z$ or other parameters, it would be an important breakthrough for
cosmological studies using SNe.

We describe in this paper a method for determining accurate distances
to SNe Ia without knowledge of redshift, based on marginalizing over
$z$.  We have used the BATM technique introduced by Tonry et
al. (2003), but the method is not specific to BATM and can be employed
with any distance estimator.  We describe the necessary steps for
implementing the method, and demonstrate that it produces accurate
distances for a large sample of Hubble flow SNe Ia.  The potential to
remove the need for spectroscopic redshifts promises to greatly
increase the number of SNe that may be used for cosmological analysis,
which is of particular importance due to the number of extremely
wide-field surveys planned for the near future.

\section{Marginalization over Redshift}
\label{batmloop-method}

As described above, all distance measurement techniques for SNe Ia
calculate the probability of a fit for a SN as a function of ($t_0$,
$A_V$, $R_D$, $z$, $d$).  The parameters are either fit as a group or
marginalized over, in order to obtain a measurement of $d$.  After
marginalizing out parameters ($t_0$, $A_V$, $R_D$), one is left with
probability as a function of ($z$, $d$).  For some questions knowledge
of ($z$, $d$) probability contours is very interesting.  The most
immediately obvious is for probing cosmological density parameters,
which led to the accelerating universe result.  For all conventional
distance determination methods, $z$ is required to be an input
parameter for the reasons described above, providing a final estimate
of $d$ by evaluating at the known redshift.

One could alternatively treat $z$ as a free parameter to be
marginalized out.  For many questions there is no added benefit to
knowing $z$ in addition to distance.  For example, inquiries into
rates as a function of distance, galaxy-type, location within the
host, or host galaxy properties can be answered based solely upon $d$,
rather than $z$.  Even with a poor constraint on $z$, in principle we
may still be able to measure accurate distances since the effect of
redshift on distance comes from dependence on (1+$z$).

Implementation of the redshift marginalization involves several minor
modifications to any fitting process.  The most important is that care
must be taken to place calculated fit probabilities for each redshift
on an equal footing, so that at each step they may be meaningfully
combined.  A suitable prior for $z$ must also be applied.  The most
obvious to consider is to scale by volume, which can be accomplished
using a uniform redshift step, weighting each by $z$.  The same prior
could be implemented with a uniform step in log $z$, weighting by
$z^2$.  A different prior to consider is one constant in $z$, rather
than volume, which could be obtained by a linear step in redshift
$without$ scaling by $z$.  One then iterates over the range of $z$
contributing non-negligible values of probability, combining the
results to produce a final estimate of $d$.

It is important to note that this redshift marginalization method is
$not$ simply a restatement of the photometric redshift procedure
typically used for galaxies.  Photometric redshifts use the observed
colors of galaxies in order to produce a likelihood range for the
redshift based on template SEDs.  Since galaxies are not standard
candles, only a very weak constraint on distance is possible by
considering magnitude.  The redshift-independent implementation of a
SN Ia distance estimator uses observed light-curves in order to
produce a constraint on both distance and redshift by considering the
magnitude and light-curve shape as well as the color.  This additional
information greatly strengthens the power of the method in comparison
to galaxy photometric redshifts.

It is also important to stress that this redshift marginalization
procedure may be implemented with $any$ method for measuring distances
to Type Ia supernovae (stretch, $\Delta m_{15}$, MLCS, BATM, or any
other equivalent method).  Rather than evaluating $d$ for the measured
spectroscopic redshift, derived probabilities should be measured for a
wide range of $z$, with $z$ then marginalized over as is done with
other fit parameters.

\section{Accurate Redshift Independent Distances}
\label{batmloop-distances}

The BATM method for calculating luminosity distances for SNe Ia was
briefly described by Tonry et al. (2003) and Barris et al. (2004),
with a more complete treatment to appear in Barris (2004).  It
is based upon an idealized set of representative SN Ia light-curves
which are photometrically and spectroscopically well-sampled in time,
and for which accurate distances are known.  With such a
spectrophotometric template set, predicted light-curves could be
produced to compare to observations.  However, data of this quality
are extremely uncommon at present, so we use a set of light-curves
which have excellent temporal coverage over a range of wavelengths and
span a wide range of luminosity, and a large set of observed spectra.
For a given redshift, the SEDs are shifted and warped so that they
match the observed photometry of each template light-curve.  BATM
treats the ``template'' and ``unknown'' in a fundamentally different
manner from the previous methods.  The SEDs and light-curves are
shifted to the redshift of the SN to be measured, so that redshift
effects are $introduced$ to the template set rather than $removed$
from the observational data.  The idea is to compare the observed SN
to what we would expect the template to look like at a given redshift,
as opposed to comparing the template to what the SN would look like
were it at the redshift of the template.

Since in BATM it is the template data that is transformed, rather than
the observed data, the operation can naturally be performed over a
range of redshift, facilitating the implementation of the redshift
marginalization.  It should be stressed again that $any$ method for
measuring distances may similarly be extended by merely iterating the
procedure over a range of redshifts, treating $z$ as an additional
free parameter and marginalizing it out.

\begin{figure}
\epsscale{0.5}
\plotone{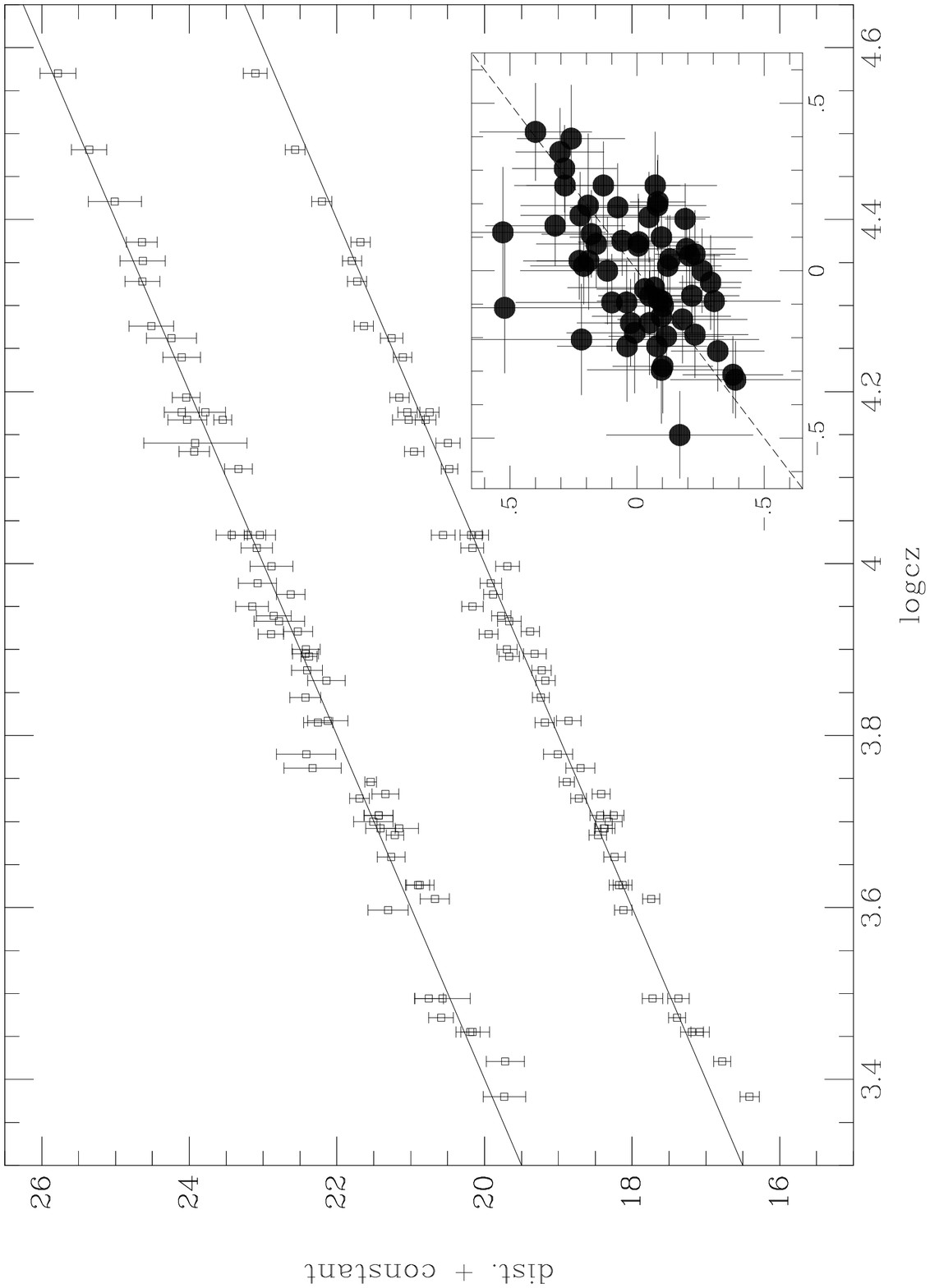}
\caption{Hubble Diagram plots for 60 nearby SNe Ia.  Bottom line shows
distances taken from Tonry et al. (2003), with a scatter of 0.19 mags.
Top values are calculated with the redshift-independent
marginalization procedure in combination with the BATM method, with a
scatter of 0.21 mags.  The x-axis values in both cases are
$spectroscopic$ redshifts, $not$ redshifts calculated from the SN
photometry.  Plots of residuals relative to the Hubble diagrams are
shown in the inset, with x-axis values for Tonry et al. (2003)
distances and y-axis from $z$-free distances.  There is a clear
correlation, indicating that the residuals are due to properties of
the SNe rather than some aspect of the BATM method or the redshift
marginalization procedure.  Dotted diagonal line is $not$ a fit to the
data, but is to guide the eye.  }
\label{hubcompare}
\end{figure}

We have performed the redshift-marginalization procedure in
combination with the BATM method for a set of 60 SNe Ia taken from
Hamuy et al. (1996), Riess et al. (1999), and Jha (2002).  This is a
well-observed sample for which we know relative distances via
spectroscopic redshifts.  Results are shown on a Hubble diagram in
Figure ~\ref{hubcompare}.  It is important to note that the redshifts
used to construct these Hubble diagrams were measured
spectroscopically, $not$ photometrically.  The RMS scatter about the
best-fit line is 0.21 mags for the $z$-free BATM distances, compared
to 0.19 mags for distances taken from the recent compilation of Tonry
et al. (2003), who calculated distances by combining the results from
as many methods as possible.  We observe no significant difference in
performance between the redshift priors mentioned in the previous
section, indicating that neither the redshift prior nor details of its
implementation are limiting factors for the method for this sample.
We have therefore used distances calculated with a linear step of
$z$=0.01 in redshift and weighted by $z$, beginning at $z$=0.01 and
typically truncated at $z$=0.20, where the probability has become
negligible.

We have scaled the uncertainties for the $z$-free BATM distances in
order to produce $\chi^{2}/N_{dof} \sim 1$ for the set of 60 SNe.  Our
initial uncertainties were overestimated in comparison to their
scatter about the Hubble line, as shown by a total $\chi^{2}$ value of
40.0 for 60 objects.  We expect that a ``training'' process for BATM
similar to that used to produce templates for other methods will both
alleviate this discrepancy and reduce the scatter about the Hubble
diagram.  Such a procedure would determine how to optimally calculate
and combine distance estimates from each of the BATM template
light-curves, and has yet to be performed.  At present it seems
sensible to rescale the uncertainties so that they more accurately
reflect the scatter and yield $\chi^{2}/N_{dof} \sim 1$.
Uncertainties in Figure ~\ref{hubcompare} reflect this rescaling.  The
distance uncertainties from Tonry et al. (2003) appear to be
$under$estimated for this set of objects, but we have left them as
reported.

\begin{figure}
\epsscale{0.5}
\plotone{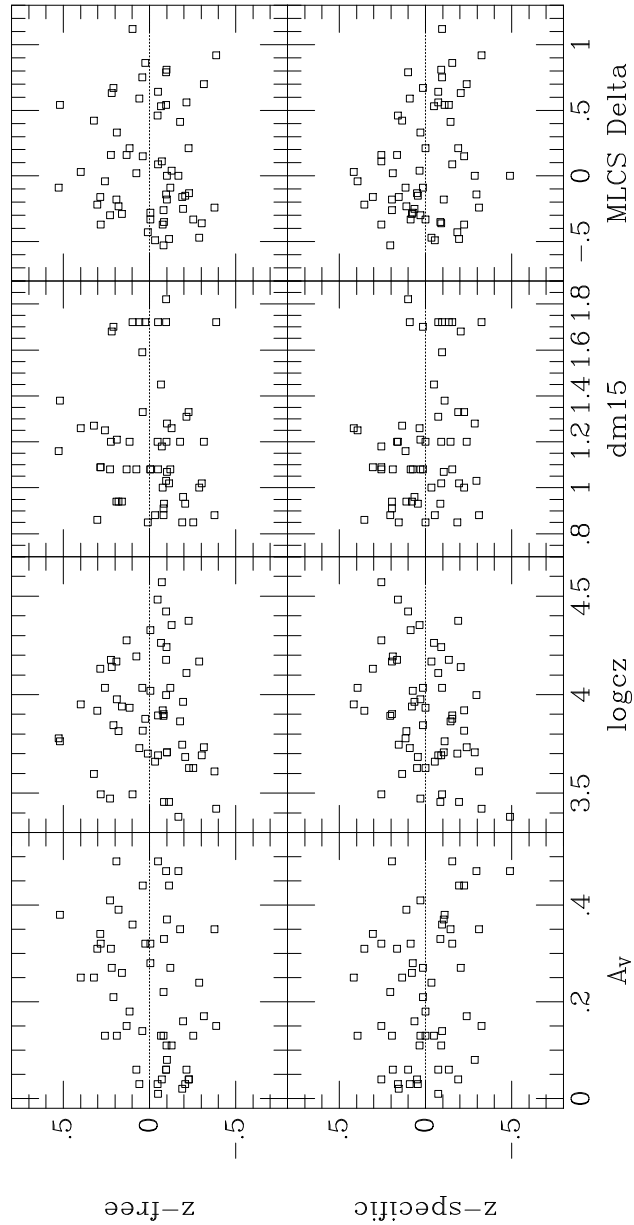}
\caption{Hubble diagram residuals as a function of SN properties for
distances calculated with the redshift-independent marginalization
procedure in combination with the BATM method (top), and for those
calculated from various redshift-specific methods (bottom).  There are
no apparent biases in the redshift-independent distances as a function
of extinction (values taken from Tonry et al. 2003), redshift, or
light-curve shape.  In fact, for large values of $A_V$, $z$, and MLCS
$\Delta$, the redshift-independent distances appear to be slightly
$less$ biased than redshift-specific methods.}
\label{residscompare}
\end{figure}

The residuals with respect to a Hubble line for the two sets of
distances are compared in the inset of Figure ~\ref{hubcompare}.
There is a clear correlation between the two methods, indicating that
the residuals are due to intrinsic properties of the SNe rather than
some feature of the redshift marginalization or the BATM analysis.
This is further evidence that the redshift marginalization procedure
is recovering the same distance information as the established
$z$-specific techniques.  It is natural to worry about additional
possible systematic biases that might be introduced by this method.
In Figure ~\ref{residscompare} we compare the Hubble diagram residuals
as a function of several SN properties.  There are no evident biases
in the redshift-independent distances as a function of extinction,
redshift, or light-curve shape.  It is interesting to note that for
SNe Ia at the high ends of the sample distribution of $A_V$, $z$, and
MLCS $\Delta$, the redshift-independent distances appear to be
slightly $less$ biased than those using redshift-specific methods.

\begin{figure}
\epsscale{0.5}
\plotone{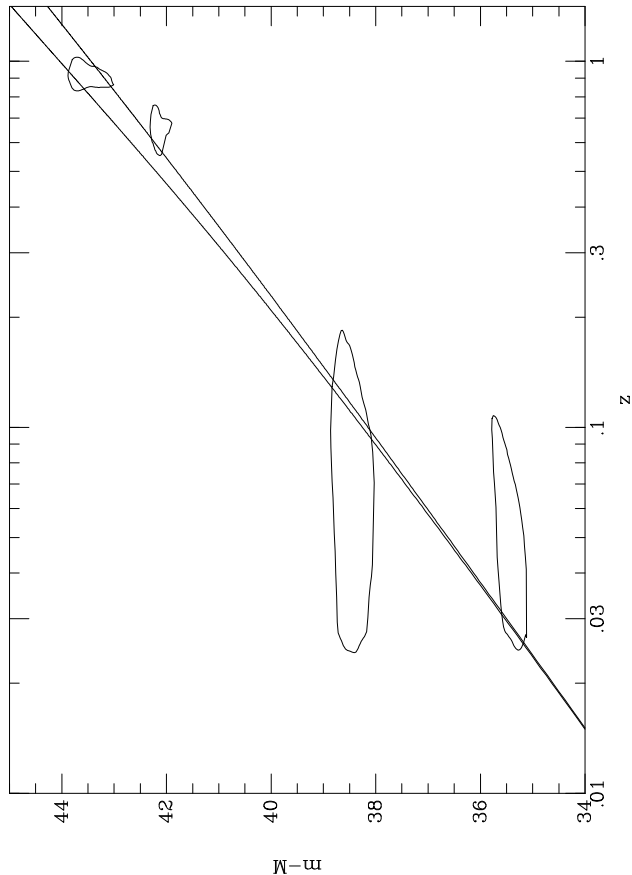}
\caption{Plot showing contours of constant probability in ($d$,$z$)
parameter space corresponding to 1-$\sigma$ for SN1996ab ($z$=0.124)
and SN1996C ($z$=0.028), from the Hubble-flow sample, as well as
SN2001iy ($z$=0.568) and SN2001jm ($z$=0.978) from Barris et
al. (2004).  Also shown are cosmological models for flat universes
with ($\Omega_{M}$,$\Omega_{\Lambda}$)=(0.3, 0.7) and (1.0, 0.0), top
and bottom, respectively.  Distance moduli are calculated with
$H_0$=72 km/s/Mpc.  Contours for the two low-$z$ SN are artificially
truncated since our minimum value for redshift was $z$=0.01.  With
further investigation, such contour plots may allow cosmological
density parameters to be determined based purely on photometric
observations of high-$z$ SNe.}
\label{dzcontours}
\end{figure}

It should in principle also be possible to examine the ($z$, $d$)
probabilities without marginalizing over $z$.  Figure
~\ref{dzcontours} shows contours in ($z$, $d$) space for two of our
Hubble flow sample as well as for two high-$z$ SN Ia from Barris et
al. (2004).  Since the $z$-free method is sensitive to (1+$z$), we get
very poor constraints on $z$ at low redshift, but for $z > 0.3$, the
constraints on $z$ and $d$ are quite similar, and good enough to begin
to differentiate between cosmological models.  Much further
investigation into the correlations between these two parameters is
necessary before the accuracy in values of ($z$, $d$) is demonstrated
as we have done for $d$ alone.

\section{Conclusions}
\label{batmloop-conclusions}

We have described a procedure for measuring accurate luminosity
distances for SNe Ia independent of knowledge of redshift by
marginalizing over $z$.  This procedure may be used in combination
with any distance measurement technique to remove the limitations on
supernova surveys imposed by the need for spectroscopic redshifts.
When applied as an extension to the BATM method, the redshift
marginalization formalism produces a Hubble diagram scatter of
$\sim$0.2 magnitudes, comparable to that using conventional
implementations of distance methods, in which knowledge of redshift is
required.  Further refinement of the BATM technique through a
``training'' process, or use with other methods, is likely to reduce
the scatter to values equal to that of redshift-specific methods.

Distances measured with this method will allow investigation of
various properties of SNe as a function of distance, which is all that
is necessary to answer many fundamental questions.  The rates of SNe
Ia as a function of redshift are still not well constrained (see Pain
et al. 1996, 2002; Tonry et al. 2003), and likely will remain so due
to the difficulty in collecting large numbers of spectroscopic
redshift.  However, rates as a function of distance are just as
fundamental, and the lack of redshift knowledge will not prevent
investigations into the dependence of rates on numerous host-galaxy
properties, for instance.  For investigations where cosmological
parameters may be considered as known, the conversion from distance to
redshift is determined, so accurate distances are all that is
necessary.

The method may also be used to produce probability contours as a
function of ($d$, $z$) for high-$z$ SNe.  This will allow cosmological
parameters to be measured based purely upon photometric observations.
Such a development will be an important step for cosmological studies
based on supernova surveys.
In order to accomplish this, it will be necessary to obtain a
sufficient number of observations to properly determine the
multi-wavelength light-curve shape to isolate the effects of ($d$,
$z$) from other fit parameters (particularly $A_V$ and $R_D$) and
sufficiently constrain the solution space.  The achievable accuracy,
and the observational strategy required to reach it, is influenced by
both the intrinsic properties of SNe Ia (which are not equally
"standard candles" at all wavelengths---see Jha 2002) and the quality
of temporal and wavelength coverage of nearby SNe Ia used for distance
calibration.

The method described here is more powerful than the photometric
redshift procedure used to estimate redshifts for galaxies based on
their observed colors, because it also takes into account the
magnitude and light-curve shape to produce constraints on both
redshift and distance.  It promises to remove the limitations to SN
studies caused by the requirement for spectroscopic redshifts for the
determination of distances for SNe Ia.  The ongoing ESSENCE (Smith et
al. 2002) and Canada-France-Hawaii Legacy Survey
(http://www.cfht.hawaii.edu/Science/CFHLS), and future projects such
as Pan-STARRS (Kaiser \& Pan-STARRS Team 2002) and the proposed
SuperNova Acceleration Probe (Nugent 2001) will also be able to
greatly increase their yields using this method, enhancing their
ability to investigate fundamental cosmological questions.

\acknowledgments

We thank the anonymous referee for numerous comments that greatly
improved the manuscript.  We thank Saurabh Jha and Brian Schmidt for
their MLCS and dm15 fits.  Financial support for this work was
provided by NASA through program GO-09118 from the Space Telescope
Science Institute, which is operated by the Association of
Universities for Research in Astronomy, Inc., under NASA contract NAS
5-26555.  Further support was provided by the National Science
Foundation through grant AST-0206329.

\clearpage

\clearpage

\clearpage

\end{document}